\documentclass[%
 reprint,
 amsmath,amssymb,
 aps,
]{revtex4-2}
\usepackage{comment}
\usepackage{graphicx}
\usepackage{float}
\usepackage{bm}
\usepackage{xcolor}
\usepackage{caption}

\begin{document}

\title{Confinement Geometry Regulates Fluctuations, Densities, and Collective Escape in Interacting Active Matter}

\author{Renu Parashar}
\author{Sandipan Dutta}%
 \email{sandipan.dutta@pilani.bits-pilani.ac.in}
\affiliation{%
 Department of Physics, Birla Institute of Technology and Science, Pilani, Rajasthan, 333031, India 
}%

\date{\today}

\begin{abstract}
Active matter, characterized by non-equilibrium signatures such as the violation of the fluctuation--dissipation theorem, giant number fluctuations, and non-Gaussian velocity distributions, is typically studied in bulk systems or trapped single-particle systems. Here, we investigate how the steady-state behavior of multiple sterically interacting Active Ornstein-Uhlenbeck Particles (AOUPs) confined within a double-well potential is influenced by the particle number and the convexity and concavity of the trapping potential. We find that increasing the particle number in our confined system leads to suppression of non-equilibrium fluctuations due to crowding unlike bulk active systems where fluctuations are enhanced at large particle numbers (Giant Number Fluctuations). This results in the steady-state density profiles of the AOUP system increasingly resembling those of passive systems for large particle numbers. However, the instantaneous density even at large particle numbers exhibits both unimodal and bimodal distributions, giving rise to large density fluctuations that are absent in passive systems, where the density remains bimodal. In the concave regions of the potential, both the velocity and single-particle position distributions are unimodal and Maxwellian, whereas they become bimodal in the non-concave regions. Confinement suppresses long-range correlations while generating localized fluctuations that depend sensitively on the potential geometry: velocity fluctuations are suppressed in the high-density concave regions and are enhanced near the convex barrier, where they exhibit pronounced non-equilibrium characteristics. At large persistence times, collective effects emerge in the form of correlated barrier crossings and strong velocity-alignment correlations, features that are absent in passive systems. These findings demonstrate how confinement geometry, particle number, and activity cooperatively regulate the spatial organization of non-equilibrium fluctuations in active matter.
\end{abstract}

\maketitle


\section{\label{sec:introduction} Introduction}

Active matter consists of nonequilibrium systems composed of self-driven units that continuously extract energy from their environment or internal reservoirs and convert it into persistent motion \cite{bechinger2016active,shaik2023confined}. This microscopic violation of detailed balance gives rise to collective behavior that has no direct equilibrium analogue. Examples span a wide range of length scales, from motor proteins, cytoskeletal filaments, bacterial colonies, and artificial microswimmers to animal groups such as fish schools, bird flocks, and ant swarms \cite{bechinger2016active,dabelow2019irreversibility}. Because activity injects energy locally at the particle scale, active systems display anomalous fluctuations, spontaneous pattern formation, rectification, clustering, flocking, and motility-induced phase separation \cite{ramaswamy2003active,marchetti2013hydrodynamics,ramaswamy2017active,mishra2010fluctuations,kuroda2023anomalous,ross2025hyperdisordered,fily2012athermal,fily2015dynamics,tailleur2008statistical,cates2015motility,solon2018generalized}. These phenomena have motivated theoretical descriptions based on hydrodynamic theories, active gel models, and scalar field theories of active fluids \cite{toner2005hydrodynamics,Kruse2004,prost2015active,solon2015pressure,Takatori2014,Wittkowski2014}.

Active matter exhibits a wide range of nonequilibrium phenomena that have no equilibrium counterpart, including persistent transport, anomalous density fluctuations, and collective self-organization \cite{bechinger2016active,marchetti2013hydrodynamics}. While many of these properties are now well understood in bulk or weakly confined systems, active particles in realistic environments rarely move in open space. Instead, they navigate through heterogeneous landscapes containing walls, pores, channels, barriers, and trapping potentials that continuously reshape their dynamics. Unlike passive Brownian particles, whose steady-state properties under confinement are determined solely by the potential through the Boltzmann distribution, active particles remain intrinsically out of equilibrium because persistent self-propulsion generates irreversible probability currents and breaks detailed balance even in static confining potentials. Consequently, confinement not only redistributes particle density but also fundamentally modifies the nonequilibrium character of the system by coupling activity to the local geometry of the potential \cite{ni2015tunable,caprini2018active,fily2015dynamics,das2018confined,malakar2018steady}. Understanding how confinement reorganizes active fluctuations is therefore essential for controlling transport, filtration, rectification, and escape processes in biological and synthetic active systems \cite{sokolov2010swimming,reichhardt2017ratchet,granek2024colloquium,Anand2024}.

A particularly useful framework for addressing these questions is a double-well potential, which naturally combines concave regions around the minima with a convex region near the barrier, allowing the influence of local curvature on active dynamics to be investigated within a single system. Previous studies have shown that a single active particle exhibits qualitatively different dynamical regimes across these regions: near the minima, the dynamics is approximately equilibrium-like, whereas close to the barrier, the persistent active force generates pronounced nonequilibrium behavior characterized by bimodal velocity distributions and enhanced irreversibility \cite{caprini2019active}. Persistent propulsion can also synchronize the escape of two interacting particles, leading to correlated barrier-crossing events \cite{caprini2021correlated}. These observations suggest that confinement can locally regulate nonequilibrium fluctuations through the interplay of activity and potential curvature. However, how these fluctuations evolve in many-body interacting systems remains largely unexplored. In particular, it is not yet understood how active density fluctuations, spatial correlations, and collective barrier-crossing dynamics are modified across the concave and convex regions of the potential, or how increasing particle interactions reshape the coexistence of equilibrium-like and nonequilibrium regions under confinement.

Here, we address this question using a one-dimensional system of sterically interacting Active Ornstein-Uhlenbeck particles (AOUPs) confined in a symmetric double-well potential. The AOUP model provides a minimal description of persistent self-propulsion through colored noise and has been widely used to study active colloids and microswimmers \cite{wittmann2017effective,fodor2016far,berthier2017new,mandal2017physical,Martin2021,Caprini2020,dutta2023most,dutta2023entropy,fily2012athermal,Farage2015,Chaki2019}. The AOUP model is particularly suitable for the present one-dimensional barrier-crossing problem because it provides a minimal continuous description of persistent self-propulsion, with the persistence time $\tau$ directly controlling the temporal correlation of the active force. In contrast, ABP dynamics is intrinsically associated with orientational diffusion in higher dimensions, while one-dimensional RTP dynamics involves
discrete reversals of the propulsion direction; AOUP therefore provides a convenient framework for relating persistence continuously to conditional
velocity statistics and barrier-crossing dynamics. The particles interact through strong short-range repulsion, which prevents particle crossing and introduces crowding. This allows us to investigate how persistence, particle density, excluded-volume interactions, and trap geometry jointly control steady-state densities, instantaneous fluctuations, single-particle statistics, velocity correlations, and collective escape events.

Our results show that the steady-state density profiles of both passive particles and AOUPs display symmetric bimodal time-averaged distributions. However, even at large persistence times, the density profiles become increasingly passive-like as the particle number increases. In contrast, the instantaneous density fields reveal a striking difference: passive systems remain nearly symmetric and weakly fluctuating, whereas AOUPs exhibit large temporal heterogeneity, asymmetric well occupation, and intermittent localization, giving rise to both unimodal and bimodal profiles.

We further show that the ER-NeR distinction becomes a collective, density-dependent property in interacting systems. Near the barrier, the particles exhibit enhanced number fluctuations (ENFs) and bimodal velocity distributions, whereas near the well bottom, the fluctuations are suppressed and the velocity distributions become unimodal and Maxwellian. This spatial partitioning is strongest for small particle numbers and large persistence times. At large particle numbers, steric crowding suppresses mobility and reduces velocity correlations.

Finally, we demonstrate that persistence and steric interactions generate correlated many-particle barrier crossings. In dilute active systems, multiple neighboring particles cross the barrier within short time intervals. These correlated escape events are absent in passive systems, where only individual barrier crossings are observed. In dense systems, correlated barrier crossings are suppressed. Our work therefore extends the single-particle ER--NeR picture and the two-particle correlated escape mechanism to many-body active systems.

Our model is also relevant to experimentally realizable confined active-colloid systems. Active Janus colloids have been studied in optical trapping potentials \cite{buttinoni2022active}, while active-particle barrier crossing has been experimentally realized using a bistable optical potential \cite{militaru2021escape}. Thus, a quasi-one-dimensional arrangement of interacting active colloids in a double-well optical potential could provide a direct experimental realization of our system, where particle tracking could probe the predicted crossover from extended correlated barrier crossing at low density to localized collective crossing under strong crowding.

The paper is organized as follows. In Section~\ref{sec:aoup}, we introduce the equations of motion for interacting AOUPs and passive particles in a double-well potential. In Section~\ref{sec:densities}, we compare steady-state and instantaneous density distributions and show that similar time-averaged densities can conceal very different nonequilibrium dynamics. In Section~\ref{sec:individual}, we analyze conditional velocity distributions and single-particle position distributions to identify equilibrium-like and nonequilibrium regions. In Section~\ref{sec:fluctuations}, we quantify enhanced number fluctuations, position and velocity fluctuations, and interparticle velocity correlations. In Section~\ref{sec:barrier crossing}, we analyze correlated barrier crossings and demonstrate how persistence promotes collective escape while crowding suppresses it.

\section{Dynamics of AOUP and passive particles in a double well potential}
\label{sec:aoup}
Consider a one-dimensional system of $N$ AOUPs with the positions of the particles $\{x_i\}_{i=1,2,...,N}$. They interact via a potential $V(x_i-x_j)$ and diffuse in an external potential $W(x_i)$. Their dynamics is governed by
\begin{align}
\dot{x}_i & = -W^{\prime}(x_i) + F_i +\eta_i.
\label{aoup}
\end{align}
$\eta_i$ is an Ornstein-Uhlenbeck process : $\langle\eta_i(t)\eta_j(t^{\prime})\rangle=\delta_{ij}v_0^2\exp(-\vert t-t^{\prime}\vert/\tau)$ with persistence time $\tau$ and variance $v_0$, also called the swim velocity. The colored noise $\eta$ satisfies: $\tau\dot{\eta}_i = -\eta_i+ v_0\sqrt{2\tau}\epsilon_i$. $\epsilon$'s satisfies $\langle\epsilon_i(t)\epsilon_j(t^{\prime})\rangle = \delta_{ij}\delta(t-t^{\prime})$. $F_i = -\frac{1}{\gamma}\partial_{x_i}\left(\sum_{j<i}V(x_i-x_j)\right)$ is a conservative force due to the interactions between the particles. $\gamma$ is the friction coefficient. In this work the external potential is a symmetrical double well potential
\begin{equation}
W(x) = \frac{W_0}{4\gamma}(x^2-a^2)^2.
\label{doublewell}
\end{equation}
The barrier is located at the origin and the two minima located at $x=\pm a$. We consider repulsive forces of strength $k$ between two particles of radius $R$ of the form
\begin{align}
F_{ij} & = k(2R-\vert x_i-x_j\vert) \text{ if } \vert x_i-x_j\vert < 2R \text{, and}\nonumber\\ & = 0 \text{ otherwise}.
\label{ineractionpotential}
\end{align}
We choose large $k$ values so that the linear force is sufficiently strong to
prevent particle crossing while avoiding excessively large forces. The linear overlap force is chosen because, in the dense systems studied here, frequent close contacts can generate very large forces for steep repulsive potentials such as the Weeks-Chandler-Andersen (WCA) potential. We expect the resulting crowding-induced suppression of mobility to remain qualitatively robust for other sufficiently steep short-range repulsive interactions, including WCA. The numerical values of the parameters used in the text are listed in Appendix \ref{appendix:parameters}.

For a system of $N$ passive particles in the double well potential of Eq \eqref{doublewell} and with repulsive interactions of Eq \eqref{ineractionpotential}, the equation of motion is
\begin{equation}
\dot{x}_i = -W^{\prime}(x_i) + F_i + \xi_i,
\label{passive}
\end{equation}
where $W$ and $F$ have the same meaning as Eq \eqref{aoup}. The noise $\xi$ satisfies $\langle\xi_i(t)\rangle = 0$ and $\langle\xi_i(t)\xi_j(t^{\prime})\rangle = 2D\delta_{ij}\delta(t-t^{\prime})$. $D$ is the thermal diffusion constant.

Hereon, the positions and lengths are scaled by the radius of the particles $R$, $x/R$ and the times by  the local relaxation time at the bottom of the traps $t_0 = \frac{\gamma}{W^{\prime\prime}(\pm a)} = \frac{\gamma}{2W_0a^2}$, $t/t_0$.

\section{Steady state densities}
\label{sec:densities}

\begin{figure*}
\centering
\includegraphics[width=\linewidth]{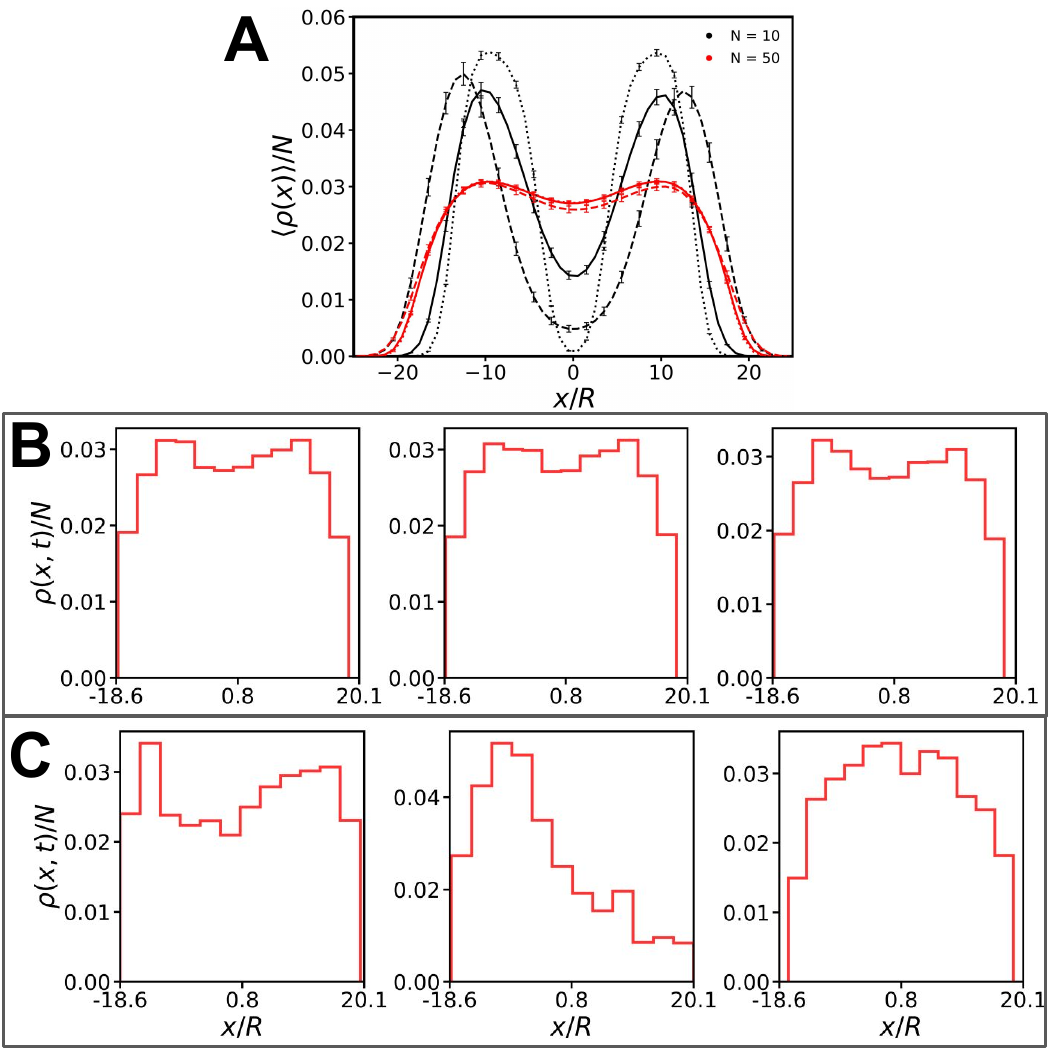}
\caption{Steady-state and instantaneous density distributions in a double-well potential. (A) Steady-state density for AOUPs with $\tau/t_0=4$ (solid), $\tau/t_0=40$ (dashed) and passive (dotted) for $N=10$, and $50$. Note that the densities are made dimensionless by multiplying by $R$. The passive systems have tiny error bars due to very small fluctuations. Increasing persistence sharpens localization within the wells, while increasing particle number broadens the density through steric crowding. (B) Representative instantaneous density profiles of 50 passive particles, showing smooth, nearly symmetric bimodal distributions. (C) Representative instantaneous density profiles of 50 active particles ($\tau/t_0=40$), exhibiting large temporal fluctuations, asymmetric well occupation, and intermittent localization due to persistent nonequilibrium dynamics.}
\label{fig:steady_state_density}
\end{figure*}

For the double-well potential, the steady-state density is obtained from simulations of Eq.~\eqref{aoup}. The simulation details and parameters are provided in Appendix-A. The mean density,
$\langle \rho(x)\rangle$, is computed by time-averaging the particle-number histograms after discarding the initial $10^4$ integration steps to ensure that the system has reached the steady state.

The resulting steady-state densities are shown in Fig.~\ref{fig:steady_state_density}-(A). In thermal equilibrium, noninteracting passive particles follow the Boltzmann distribution,
$\rho(x)\propto \exp[-\beta W(x)]$, giving rise to the familiar bimodal density centered at the minima of the double-well potential. Repulsive interactions broaden these peaks and increase the particle population near the barrier through excluded-volume effects \cite{evans1979nature,lutsko2010recent}. This trend is clearly observed for the passive system, where increasing the particle number progressively smooths the density profile and fills the barrier region.

The active system also exhibits a bimodal steady-state density for both persistence times considered ($\tau/t_0=4$ and $\tau/t_0=40$). Increasing the persistence enhances localization within the wells, producing sharper peaks and a stronger depletion around the barrier. However, increasing the particle number counteracts this localization through steric crowding, broadening the density and increasing the occupation of the central region. For sufficiently large systems ($N=50$), the density profiles for $\tau/t_0=4$ and $\tau/t_0=40$ closely resemble the passive distribution, indicating that strong excluded-volume interactions effectively suppress the influence of activity on the time-averaged density. Thus, while persistence promotes localization within the wells, crowding progressively restores an equilibrium-like spatial organization. This crossover can be qualitatively understood from the geometric length scales of the system. The particle diameter is $2R=2$, whereas the two potential minima are separated by a distance of $20$. For $N=10$, approximately five particles occupy each well, corresponding to a nominal excluded-volume length $5(2R)\simeq 10$, leaving sufficient configurational freedom for persistent propulsion to generate large particle displacements and collective motion. In contrast, for $N=50$, approximately 25 particles occupy each side, corresponding to a nominal excluded-volume length $25(2R)\simeq50$. Since particles cannot cross each other, frequent steric contacts strongly constrain their motion and suppress the propagation of persistence-induced fluctuations through the well. Thus, the reduced nonequilibrium signatures at large $N$ originate from crowding rather than from a reduction of the intrinsic activity; localized collective motion can nevertheless survive near the barrier.

Although the passive and active systems exhibit qualitatively similar bimodal steady-state densities, their instantaneous density fields,
$\rho(x,t)$, reveal strikingly different dynamics, as shown in Fig.~\ref{fig:steady_state_density}-(B) and (C). For passive particles (Fig.~\ref{fig:steady_state_density}-(B)), the density remains nearly symmetric between the two wells at all times, with only weak, spatially uncorrelated fluctuations arising from thermal diffusion. Consequently, the instantaneous profiles are relatively smooth and fluctuate only weakly about the steady-state distribution. In contrast, the instantaneous densities of AOUPs (Fig.~\ref{fig:steady_state_density}-(C)) exhibit pronounced temporal and spatial heterogeneity. Persistent self-propulsion generates large correlated fluctuations, producing irregular density profiles that continuously evolve between nearly symmetric bimodal configurations and strongly asymmetric states in which one well contains a substantially larger fraction of particles than the other. As a result, the instantaneous density may transiently appear almost unimodal despite the underlying bimodal steady state. These intermittent fluctuations reflect the nonequilibrium nature of active matter, where persistence induces correlated particle motion, transient clustering, and collective migration across the barrier. Therefore, while steady-state averaging masks many signatures of activity, the instantaneous density clearly distinguishes equilibrium passive dynamics from the strongly fluctuating nonequilibrium behavior of active particles.

Importantly, the increasingly passive-like steady-state density at large
particle numbers does not imply a complete disappearance of nonequilibrium collective dynamics. Even for $N=50$, the instantaneous density profiles exhibit intermittent redistribution of particles between the two wells, indicating that collective inter-well motion persists despite strong crowding. At large persistence, the active force remains correlated over the persistence time $\tau$, allowing neighboring particles near the barrier to move cooperatively. Since steric interactions prevent particle crossing, persistently driven particles can push neighboring particles toward and across the barrier. At large $N$, crowding suppresses the
propagation of this cooperative motion through the entire well, but a small group of particles close to the barrier can still undergo collective
transitions. Thus, increasing particle number produces an equilibrium-like crossover in the time-averaged density without completely eliminating localized nonequilibrium collective escape, as demonstrated more explicitly by the velocity correlations and barrier-crossing statistics discussed in Secs. \ref{sec:fluctuations} and \ref{sec:barrier crossing}.

\begin{figure*}
 \includegraphics[scale=0.7]{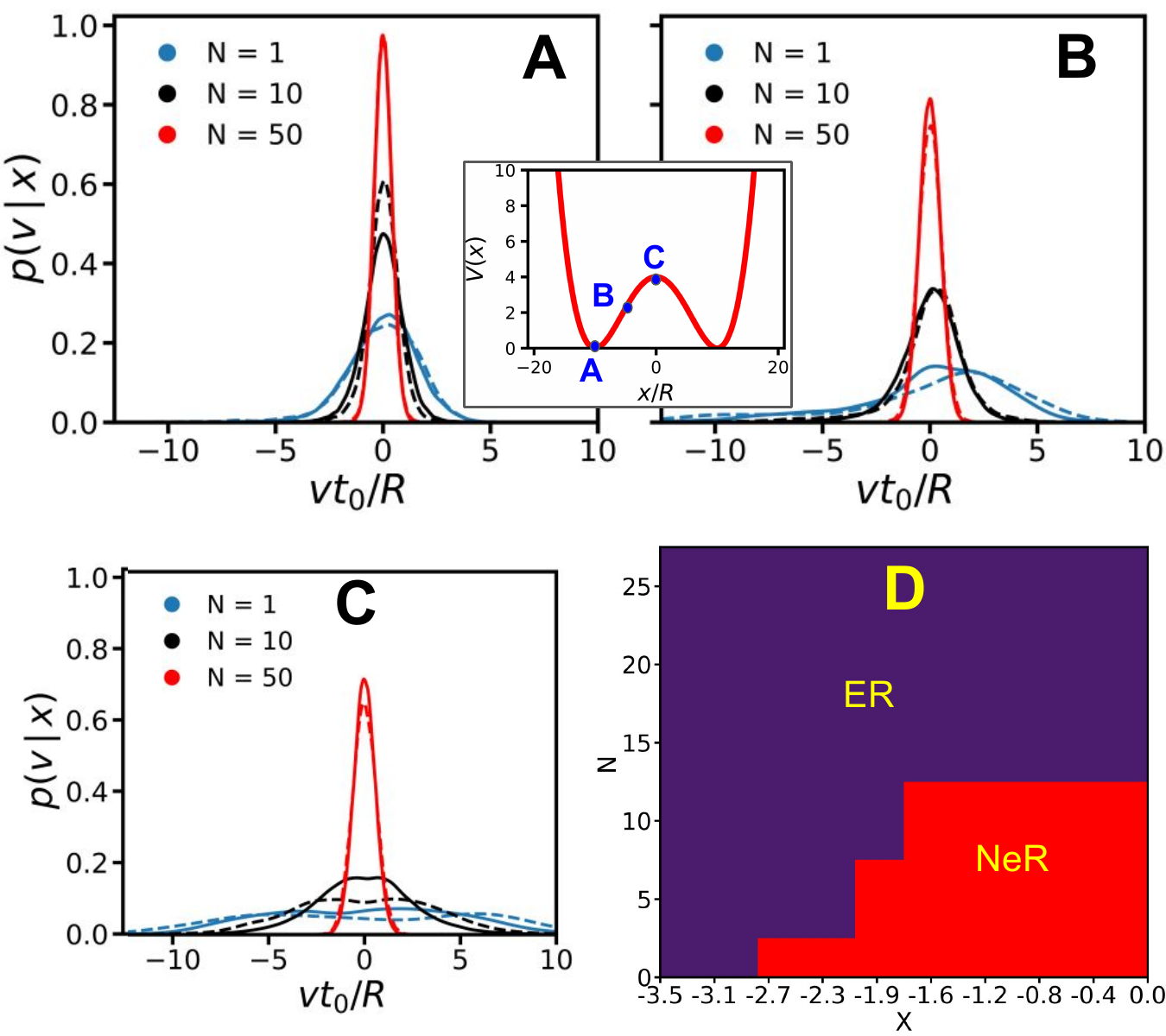}
 \caption{Conditional velocity distributions $p(v|x)$ of particles in a double-well potential evaluated at different spatial locations in the left well. Note that $p(v|x)$ is made dimensionless by multiplying by $R/t_0$. (A) At the bottom of the left well ($x/R \approx -12.5$), (B) at an intermediate position between the minimum and the barrier ($x/R \approx -6$), and (C) at the barrier top (x $\approx 0$) as shown in the inset. The distributions show a transition from unimodal Maxwell-Boltzmann-like to bimodal shapes as the position approaches the barrier. The solid (dashed) curves represent $\tau/t_0=4$ ($\tau/t_0=40$). (D) The corresponding bimodality measure as a function of position in the left well, quantified using a two-component Gaussian mixture model \cite{fraley2002model}, where the preference for a bimodal over a unimodal fit is determined via the Bayesian Information Criterion (BIC) detailed in the text. The onset of bimodality in $p(v|x)$, identified by the BIC favoring the two-Gaussian model, marks the crossover from the equilibrium-like region (ER) to the non-equilibrium region (NeR). We use $\tau/t_0=40$.}
 \label{fig:velocity_distribution}
 \end{figure*}

\section{Dynamics of Individual Particles}
 \label{sec:individual}
To understand the origins of the diminished nonequilibrium behavior at high particle numbers we investigate the dynamics of the particles in different regions of the confinement. Since the particles in this system are subject to a non-crossing constraint, the initial ordering is strictly maintained throughout the duration of the simulations. We calculate $p(v|x = x_0)$ by measuring the velocity of the particles at some fixed positions $x_0$ in the trap. Caprini \textit{et al.}~\cite{caprini2019active} showed that for a single AOUP in a double well potential, the conditional velocity distribution, $p(v|x) = p(v,x)/\rho(x)$, can be either unimodal or bimodal depending on the particle’s position within the trap. Near the potential minima, the distribution is unimodal and well approximated by a Maxwell–Boltzmann Gaussian form of velocity with an effective temperature. These regions were termed Equilibrium-like Regions (ER). In contrast, Nonequilibrium Regions (NeR), occurring near the barrier, were identified as regions where the distributions become bimodal or develop non-Gaussian tails. 

 Our analysis of $p(v|x)$, shown in Fig.~\ref{fig:velocity_distribution}-(A–C), shows the variation of the velocity distributions within the left well. The behavior in the right well is symmetric and is therefore not shown. Near the bottom of the well (A), $p(v|x)$ is unimodal and approximately equilibrium Gaussian, centered around zero velocity. Moving to intermediate positions (B), the distribution broadens and begins to deviate from a simple Gaussian form. Close to the barrier (C), $p(v|x)$ becomes distinctly bimodal, with two peaks corresponding to particles moving in opposite directions. For small particle numbers, this progression from unimodal to bimodal behavior is clearly visible as one approaches the barrier. In contrast, at larger particle numbers, steric constraints prevent particles from crossing each other, leading to accumulation near the well bottom. As a result, the velocity distributions remain unimodal and centered around zero even at positions where bimodality would otherwise emerge. Following Caprini et. al. we define the non-equilibrium region (NeR) as the spatial region where $p(v|x)$ exhibits a bimodal structure, while the equilibrium-like region (ER) corresponds to unimodal distributions. Using this criterion, we identify the spatial extent of the NeR and summarize it in the phase diagram shown in Fig.~\ref{fig:velocity_distribution}-(D). The onset of bimodality is quantified by fitting the conditional velocity distribution $p(v\mid x)$ using Gaussian mixture models containing one ($m=1$) and two ($m=2$) components. Model selection is performed using the Bayesian Information Criterion (BIC) \cite{schwarz1978estimating}, $\text{BIC} = k\ln n - 2\ln\hat{\mathcal{L}}$, where $k$ is the number of fitted parameters, $n$ is the number of observations, and $\hat{\mathcal{L}}$ is the maximum likelihood. The model with the smaller BIC is preferred because the criterion penalizes the additional parameters of the more complex model. We define $\triangle\text{BIC}=\text{BIC}(m=1)-\text{BIC}(m=2)$. Following conventional criteria for model comparison \cite{kass1995bayes, fraley2002model}, $\triangle\text{BIC}>10$ is taken as strong evidence in favor of the two-component model. Such distributions are classified as bimodal and the corresponding spatial locations as nonequilibrium regions (NeR). Gaussian-mixture-based model selection using BIC is well established for identifying the appropriate number of mixture components \cite{fraley2002model}. The distribution is identified as bimodal, indicating the presence of two underlying velocity populations. Following Caprini \textit{et al.}~\cite{caprini2019active}, regions near
the potential minima, where the conditional velocity distribution
$p(v|x)$ is approximately Gaussian, are identified as equilibrium-like
regions (ER). In contrast, persistence produces significant deviations
from Gaussian velocity statistics, particularly bimodal distributions
near the barrier, which signal intrinsically nonequilibrium dynamics
and define the nonequilibrium regions (NeR). Regions where the two-component fit is favored are classified as NeR, marking the transition from equilibrium-like to non-equilibrium behavior within the trap. The most probable velocities can be inferred from the peaks of the distributions. At low particle numbers (e.g., $N=10$), the distribution near the well bottom peaks at zero velocity (A), while at intermediate positions (B) the peak shifts, indicating directed motion. At the barrier (C), the bimodal structure reflects the coexistence of particles moving toward opposite wells, resulting in symmetric peaks about zero. For larger systems (e.g., $N=50$), the distributions remain symmetric and unimodal with a peak at zero velocity across all positions, consistent with constrained motion due to crowding.

\begin{figure*}[!htbp]
 \includegraphics[scale=0.6]{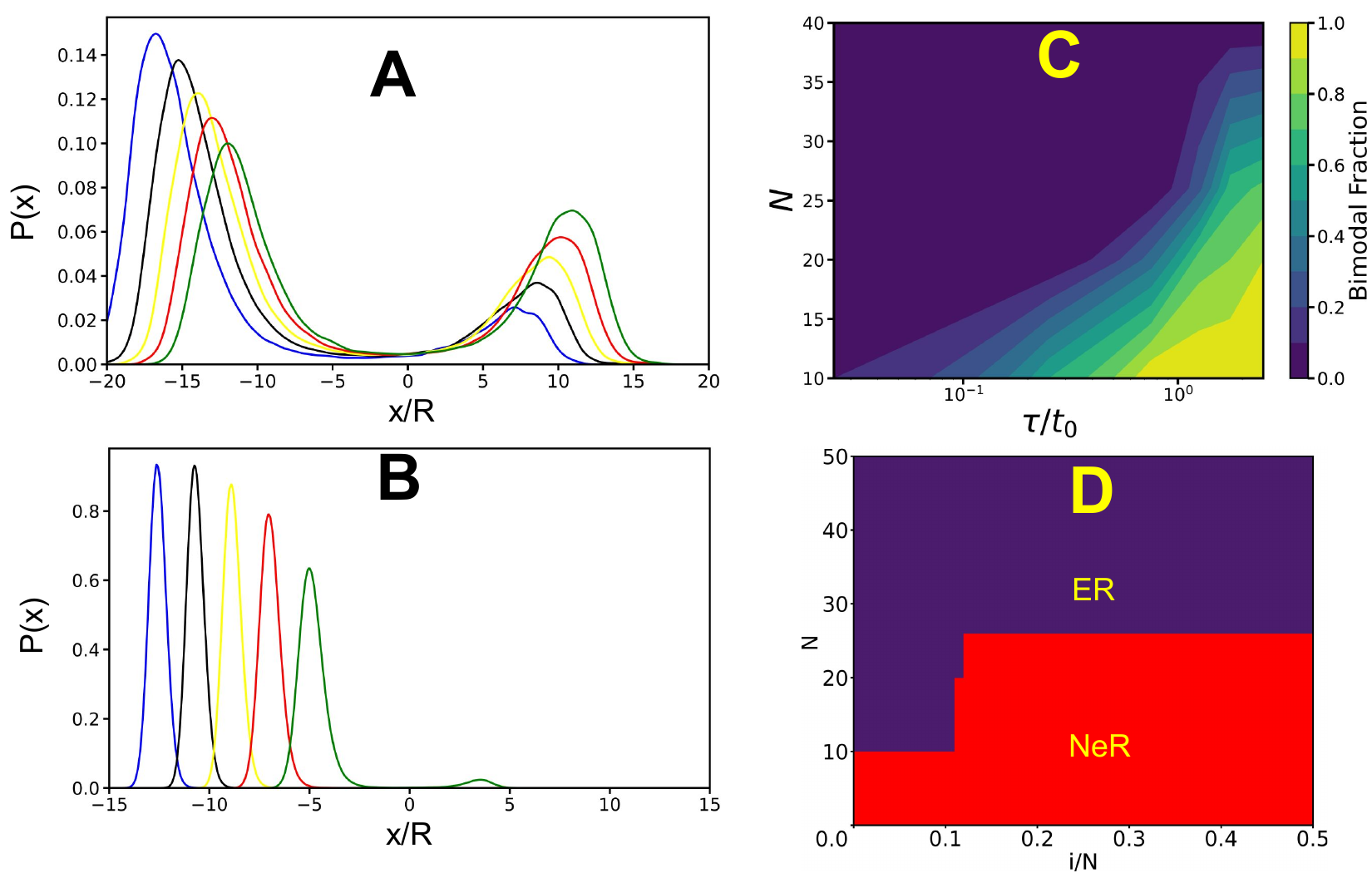}
 \caption{Statistics of individual particle position distributions in the left well of a double-well potential in steady state for a system of ten particles. The particle distributions are multiplied by $R$ to make them dimensionless. (A) Single-particle position distribution functions for five active Ornstein–Uhlenbeck particles (AOUPs) in the left well, showing that some particles develop bimodal spatial distributions due to persistent activity. (B) Corresponding single-particle position distributions for five passive particles in the left well, which remain unimodal and consistent with equilibrium fluctuations. (C) Bimodal fraction of the single-particle position distributions for AOUPs. (D) Spatial demarcation of the non-equilibrium region (NeR) and equilibrium-like region (ER) based on the bimodality (or unimodality) of the individual particle distributions as a function of particle index. The particle index represents position within the left well, where the leftmost particle has index $0$ and the particle closest to the barrier corresponds to 
$0.5$. We use $\tau/t_0=40$.}
 \label{fig:individual_particles}
 \end{figure*}

While the steady-state densities of active and passive particles appear similar, our simulations reveal distinct behavior at the level of individual particles. Since the particles in this system can not cross each other, the initial ordering is strictly maintained throughout the duration of the simulations. Consequently, each particle index serves as a reliable proxy for its spatial location within the potential. For a system of $10$ particles equally distributed within the trap, the $3$rd particle remains localized near the bottom of the trap, while the $5$th particle stays close to the origin. In the $50$-particle case, the $13$th and $25$th particles are situated near the potential minimum and the barrier, respectively. We bin the trajectory of each particle into a histogram to calculate the single particle position distribution, $P_i(x) = \langle \delta(x_i(t)-x)\rangle$. If we sum over the single position distribution of all $N$ particles we recover the steady state densities of Fig \ref{fig:steady_state_density}.  Since the trajectories do not intersect, we obtain  the single particle distribution for each particle in case of $10$ particles in Fig.~\ref{fig:individual_particles}-(A) (only particles in the left well are shown) that have two distinct peaks in each of the wells. The corresponding plot for passive particles in Fig.~\ref{fig:individual_particles}-(B) displays single peaks. Active particles undergo frequent inter-well hopping, whereas the passive particles remain localized within a single well, yielding unimodal distributions. If we increase the number of active particles the single particle distributions centered near the well bottoms become increasingly unimodal as crowding leads to less frequent barrier crossings, whereas those near the barrier stay bimodal. We calculate what fraction of particles have bimodal single particle position distributions as we vary the persistence time $\tau$ and the total particle number $N$ as seen in Fig.~\ref{fig:individual_particles}-(B). As expected, low activity and particle numbers favor bimodality of the single particle position distribution. Now we define ER and NeR based on the bimodality of the single particle position distributions. Particles situated within NeRs and possessing higher activity are more likely to overcome the potential barrier, resulting in bimodal position distributions. Conversely, increasing $N$ leads to crowding, reduced mobility, and enhanced confinement within a single well, thereby restoring unimodality. We find a qualitative agreement of the ERs and NeRs obtained from the velocity distributions and individual position distributions of Fig. \ref{fig:velocity_distribution}-(D) and Fig. \ref{fig:individual_particles}-(B) respectively. At large $\tau$ most particles have bimodal (unimodal) distribution (Fig. \ref{fig:individual_particles}-(C)) for small (large) $N$. 

\section{Fluctuations and correlations}
\label{sec:fluctuations}

 \begin{figure*}
 \centering
 \includegraphics[scale=0.7]{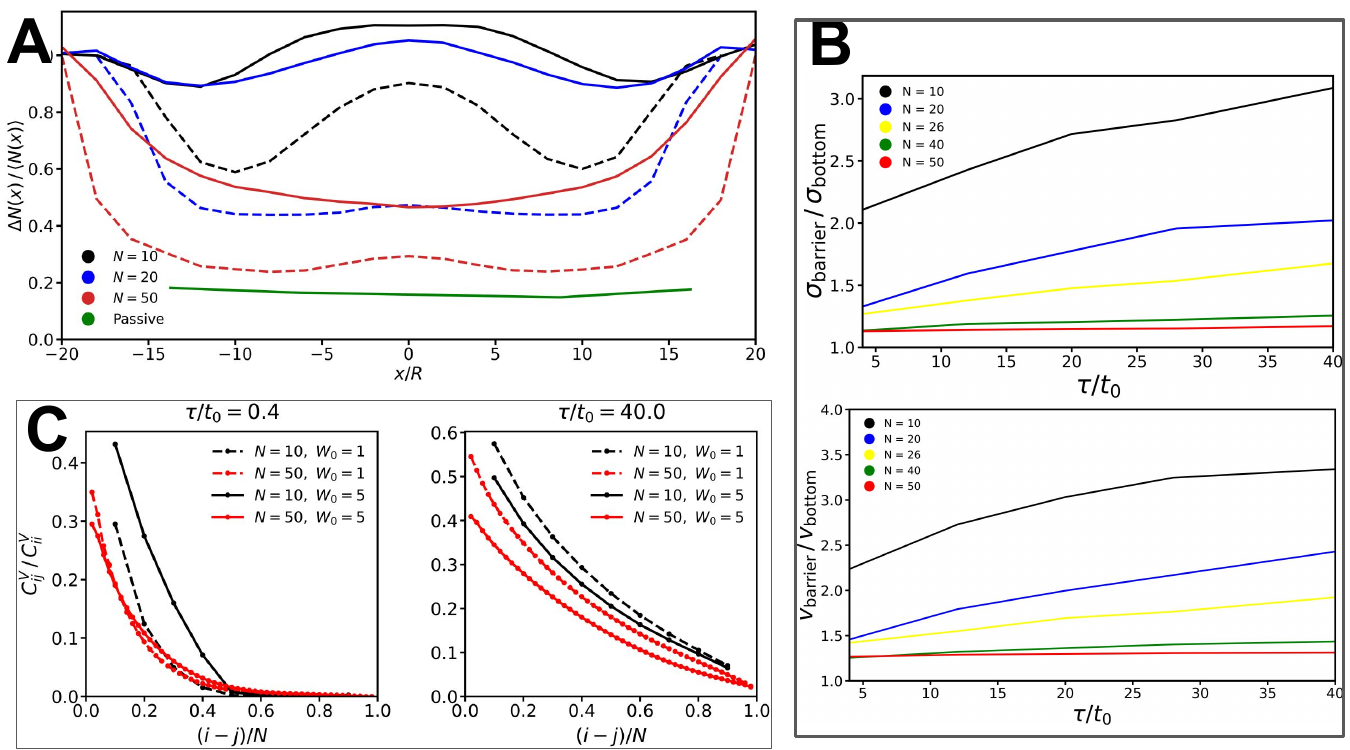}
 \caption{Fluctuations and correlations of AOUPs in a double-well trap. (A) Enhanced Number Fluctuations as a function of position for different particle numbers. Results are shown for $\tau/t_0=4$ (dashed) and $\tau/t_0=40$ (solid); active particles exhibit the largest fluctuations in the nonequilibrium regions (NeR) near the barrier and the smallest in the equilibrium regions (ER) near the well minima. (B) Ratios of the position fluctuations, $\sigma_{\rm barrier}/\sigma_{\rm bottom}$ (top), and rms velocity fluctuations, $v_{\rm barrier}/v_{\rm bottom}$ (bottom), demonstrating the enhanced fluctuations near the barrier with increasing persistence. (C) Normalized average interparticle velocity correlations, $C_{ij}^{v}/C_{ii}^{v}$, as a function of normalized particle separation $(i-j)/N$ for $N=10$ and $N=50$ at $\tau/t_0=0.4$ and $\tau/t_0=40$, showing that activity enhances the spatial range of velocity correlations, while increasing particle number suppresses them.}
 \label{fig:fluctuations_correlations}
 \end{figure*}

Nonequilibrium fluctuations in active systems play the key role in generating directed motion in ratchets, even without net external forces \cite{granek2024colloquium}. Unlike thermal fluctuations in equilibrium Brownian ratchets, these fluctuations break detailed balance and allow asymmetric substrates or barriers to rectify motion like in bacterial baths \cite{sokolov2010swimming, reichhardt2017ratchet}. ENF is quantified by the number fluctuations scaled by the square root of the average number of particles, i.e. $\triangle N(x)/\sqrt{\langle N(x)\rangle}$ for active systems. In our system, we obtain the local fluctuations and average particle number at a given position from the standard deviation and mean of particle number respectively in a small window about that position in Fig \ref{fig:fluctuations_correlations}-(A). $\triangle N(x)/\sqrt{\langle N(x)\rangle}$ increases with increasing activity and decreases with large $N$. Fluctuations are suppressed at the well bottom and enhanced at the barrier for small $N$. At large $N$ the fluctuations are almost uniform like the passive systems. The long ranged ENF for bulk systems becomes localized in confined active systems. The single particle position distributions in Fig \ref{fig:individual_particles}-(A) are narrowest for the active particles at the well bottom, hence have the lowest fluctuations. The distributions are broadest near the barrier with highest fluctuations. To quantify the spatial variation of density fluctuations, we consider the ratio of the position fluctuations of the particle near the barrier, $\sigma_{\rm barrier}$, to those near the bottom of the well, $\sigma_{\rm bottom}$. As shown in Fig.~\ref{fig:fluctuations_correlations}-(B, top), the ratio $\sigma_{\rm barrier}/\sigma_{\rm bottom}$ is significantly larger than unity for small particle numbers and increases with persistence time, demonstrating that fluctuations are strongly amplified in the nonequilibrium regions (NeR) around the barrier. Persistent active motion enables particles to repeatedly explore the barrier region, producing enhanced positional fluctuations and density inhomogeneities. As the particle number increases, steric interactions and crowding suppress these fluctuations, causing the ratio to approach unity. For $N=50$, the fluctuations become nearly uniform throughout the trap and are essentially independent of persistence time, resembling the behavior of the passive system. These results show that crowding progressively suppresses activity-induced fluctuations and drives a crossover from a fluctuation-dominated nonequilibrium regime near the barrier to an equilibrium-like regime near the minima, where confinement and particle interactions effectively homogenize the dynamics.

The same trend is observed for the velocity fluctuations. The ratio of the root-mean-square velocities at the barrier and the well bottom,
$v_{\rm barrier}/v_{\rm bottom}$, shown in Fig.~\ref{fig:fluctuations_correlations}(B, bottom), increases strongly with persistence for dilute systems, indicating enhanced velocity fluctuations in the NeR. In contrast, the ratio remains close to unity for $N=50$, demonstrating that crowding suppresses the nonequilibrium enhancement of particle velocities.

To characterize the collective dynamics generated by persistence, we compute the average interparticle velocity correlations shown in Fig.~\ref{fig:fluctuations_correlations}-(C). In this one-dimensional system, each particle velocity is mapped onto its instantaneous direction, $\mathrm{sgn}(v_i)=\pm1$, and the correlation function is defined as
\[
C_{ij}^{v}=\left\langle \mathrm{sgn}(v_i(t))\,\mathrm{sgn}(v_j(t))\right\rangle .
\]
For weak activity ($\tau/t_0=0.4$), the correlations decay rapidly with particle separation, indicating predominantly local and weakly correlated motion. Increasing the persistence to $\tau/t_0=40$ substantially extends the spatial range of the correlations, reflecting the tendency of neighboring particles to move coherently over long distances. The effect is particularly pronounced for $N=10$, where correlations persist across almost the entire system, demonstrating strong collective motion. As the particle number increases to $N=50$, frequent collisions disrupt coherent motion, leading to a faster decay of the correlations and a more localized collective response. Thus, persistence promotes long-ranged cooperative dynamics, whereas steric crowding suppresses velocity correlations and drives the system toward equilibrium-like behavior. The emergence of long-ranged velocity correlations at large persistence provides a microscopic explanation for the enhanced ENFs and the increased probability of simultaneous barrier-crossing events observed in dilute active systems.

\section{Correlated barrier crossing}

\begin{figure*}
 \includegraphics[scale=0.68]{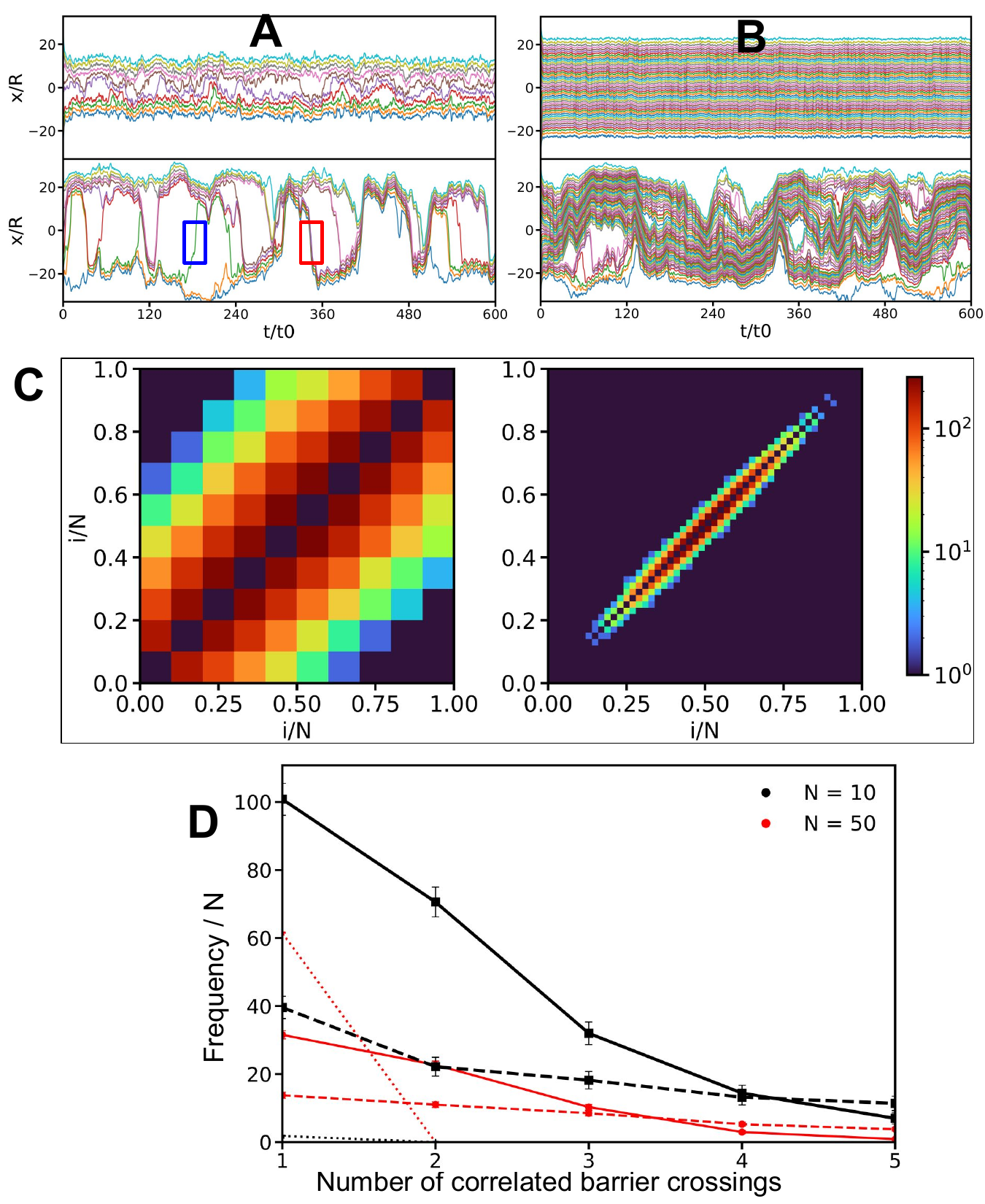}
 \setlength{\abovecaptionskip}{0pt}
 \caption{Stochastic Dynamics and Collective Barrier Crossing. (A) Representative trajectories $x(t)$ for a system of $N = 10$ particles. The top panel shows the weak persistence regime ($\tau/t_0=0.4$), where particles fluctuate independently within the wells. The bottom panel depicts the high persistence regime ($\tau/t_0=40$), where particles exhibit correlated transitions (shown in red box) between the potential minima. The double }
 \label{fig:barrier_crossing}
 \end{figure*}

 \begin{figure*}[t] 
    \ContinuedFloat
    \centering
    \caption{ (Continued) well potential is also shown on the left. The blue and red box depict the single-particle and multiparticle barrier crossings respectively. (B) Trajectories for a larger system of $N = 50$ particles under the same persistence conditions ($\tau/t_0=0.4$ top, $\tau/t_0=40$ bottom), highlighting the scaling of collective behavior with increased particle density. (C) Heatmaps representing  the correlations of jumps of individual particles  between wells for $N = 10$ and $50$ particles. The narrowing of the diagonal band indicates the spatial localization of correlated transition events as $N$ increases. 
 (D) Normalized frequency of correlated barrier-crossing events as a function of the number of particles participating in each event. The solid, dashed and dotted curves denote $\tau/t_0=4$, $\tau/t_0=40$, and passive systems respectively. A correlated event is defined as two or more particles crossing the barrier within a time window $\Delta t_c/t_0 = 16$; exact coincidence of the crossing times is not required. The same $\Delta t_c$ is used for the active and passive systems. Active systems exhibit a substantially larger probability of multi-particle correlated crossings than the passive system, with the strongest synchronization occurring for $N=10$.}
 \end{figure*}

\label{sec:barrier crossing}
The particle trajectories in Fig.~\ref{fig:barrier_crossing}-(A,B) reveal that both activity and particle number govern the collective barrier-crossing dynamics. For the dilute system ($N$ = 10), particles undergo frequent transitions between the two wells. At small persistence ($\tau/t_0$ = 0.4), barrier crossings occur independently and diffusively, whereas at large persistence ($\tau/t_0$ = 40) several neighboring particles cross the barrier within a short time interval, producing pronounced collective switching events highlighted by the colored boxes. These correlated crossings originate from the finite persistence of the active force together with steric repulsion: when particles approach the barrier, those behind continue to exert a persistent propulsive force on the leading particles, effectively synchronizing their escape and generating near-simultaneous barrier-crossing events that are absent in passive systems \cite{caprini2021correlated}. In contrast, for the dense system ($N$ = 50) the trajectories remain strongly localized within their respective wells even at large persistence, with only particles close to the barrier participating in occasional collective transitions while particles deeper inside each well remain trapped by crowding.

To quantify the collective barrier-crossing dynamics, a crossing event for particle $i$ is identified when its position changes from $x_i<0$ to $x_i>0$, or vice versa. We use the term "correlated barrier crossing" rather than requiring particles to cross the barrier at exactly the same instant. Two or more particle crossings are classified as a correlated multi-particle event when they occur within the same finite time window $\Delta t_c/t_0 = 16$. The same time-window criterion is applied to both active and passive systems and is chosen to capture correlated crossing events while avoiding double counting due to rapid recrossings. The correlation matrices shown in Fig.~\ref{fig:barrier_crossing}-(C) demonstrate that the extent of collective motion depends strongly on particle number. For $N$ = 10, the broad band of large values indicates that almost all particles participate in correlated barrier-crossing events, reflecting system-wide synchronization. As the particle number increases, synchronization becomes increasingly localized: only particles near the barrier ($i$/$N \approx$ 0.5) exhibit strong correlations, while particles farther from the barrier become essentially uncorrelated. For $N$ = 50, the narrow diagonal band shows that synchronization is restricted to a few neighboring particles close to the barrier, whereas particles beyond the well bottoms rarely participate in collective escape. This demonstrates that crowding suppresses the spatial extent of cooperative motion.

The statistics of correlated barrier crossings are summarized in Fig.~\ref{fig:barrier_crossing}-(D). For active systems with $N=10$, events involving two or more particles crossing the barrier within the same time window occur with high probability, confirming the emergence of long-range synchronization throughout the dilute system. This collective escape originates from the interplay of persistent self-propulsion and steric interactions. As particles approach the barrier, the active force remains correlated over a time scale $\tau$, so trailing particles continue to exert a persistent propulsive force on those ahead. Because particle crossings are prohibited by the steric repulsion, this sustained pushing causes neighboring particles to move as a correlated cluster. Once the leading particle overcomes the barrier, the particles behind, already aligned and persistently driven in the same direction, rapidly follow, producing near-simultaneous barrier-crossing events. Consequently, increasing the persistence time enhances both the probability and the size of these collective escape events, as reflected by the higher frequencies of multiple-particle crossings in Fig.~\ref{fig:barrier_crossing}-(D).

Increasing the particle number progressively suppresses these events, and for $N=50$ the probability decays rapidly with the number of simultaneously crossing particles, indicating that collective escape becomes confined to only a small cluster near the barrier. At high densities, steric crowding strongly restricts particle mobility and inhibits the propagation of persistent active forces through the system. As a result, only particles located close to the barrier can participate in a cooperative escape event, while particles deeper inside the wells remain effectively caged by their neighbors.

In the passive system, barrier-crossing events are predominantly single-particle events, with a much smaller occurrence of correlated multi-particle crossings than in the active system. Since the passive particles interact through steric repulsion, their barrier crossings need not be strictly statistically independent. However, in the absence of a persistent active force, thermal fluctuations cannot sustain the cooperative pushing mechanism that produces the pronounced correlated multi-particle escape observed for AOUPs. Thus, persistence strongly enhances the temporal correlation between neighboring barrier-crossing events, leading to correlated collective escape. These results therefore generalize the two-particle correlated escape mechanism reported by Caprini \textit{et al.}~\cite{caprini2021correlated} to interacting many-body systems, showing that persistent activity generates collective barrier crossing in dilute active matter, whereas steric crowding progressively restores equilibrium-like, independent barrier-crossing dynamics as the particle density increases.

\section{Conclusions}
\label{sec:conclusions}

In this work, we have systematically examined how confinement, activity, and particle number influence the steady-state and dynamical properties of interacting Active Ornstein–Uhlenbeck Particles (AOUPs) in a double-well potential, and compared them with their passive counterparts. Through  numerical simulations, we identified clear signatures of nonequilibrium behavior that arise exclusively in active systems and traced how these features evolve as the number of particles and their persistence time vary.

Although the steady-state density profiles of active and passive systems appear qualitatively similar and are both bimodal due to the underlying double-well potential, their instantaneous density distributions reveal markedly different behaviors. Passive systems remain confined within their respective wells and display stable bimodal distributions at all times, reflecting the Boltzmann-like equilibrium state. In contrast, AOUPs exhibit rich transient structures, including unimodal and bimodal density distributions, due to occasional barrier-crossing events and clustering driven by persistent propulsion. These short-lived unimodal states occur when small groups of particles collectively jump between wells, reflecting the inherently cooperative nature of active fluctuations under confinement - features that are not seen in passive systems.

The analysis of velocity and position distributions provided a clear framework to distinguish ER and NeR within the trap. Near the potential minima, both the conditional velocity distributions and single-particle position distributions are unimodal and approximately Maxwellian, signifying effective thermalization. Near the barrier, however, these distributions become bimodal and strongly non-Gaussian, marking the emergence of NeR. The extent of these regions depends sensitively on the persistence time $\tau$ and particle number $N$. For small $\tau$ or large $N$, crowding and steric hindrance suppress fluctuations and diminish bimodality, restoring equilibrium-like behavior. For large $\tau$ and small $N$, persistence amplifies fluctuations and promotes barrier-crossing, driving the system deeper into nonequilibrium regimes.

By quantifying the ratio of fluctuations and the bimodality coefficient, we showed a clear crossover from fluctuation-dominated NeR at the non-concave regions of the potential to fluctuation-suppressed ER at concave regions. This transition illustrates how the geometry of confinement and particle crowding can serve as control parameters to regulate nonequilibrium fluctuations in active matter. Furthermore, velocity–velocity correlations revealed strong, longer-ranged coherence in small systems, signifying collective motion, while in larger systems, correlations became increasingly local and short-ranged due to frequent reorientations and averaging over many interactions. This transition from longer-range coherence to local correlations marks a suppression of enhanced number fluctuations and an approach toward mean-field-like behavior as particle number increases.

The barrier-crossing analysis demonstrates that correlated escape is a genuine many-body nonequilibrium phenomenon arising from the interplay of persistence, steric interactions, and confinement. While persistent active forces promote correlated motion by allowing trailing particles to continuously propel those ahead, increasing particle density limits the propagation of these correlations through crowding. Consequently, dilute systems exhibit system-wide, long-range synchronization with frequent multi-particle barrier-crossing events, whereas dense systems support only localized cooperative escape involving particles close to the barrier, with the remaining particles effectively immobilized within the wells. In the passive system, barrier crossings are predominantly isolated single-particle events, whereas persistent activity strongly enhances correlated multi-particle escape, confirming that the pronounced synchronization observed here is an intrinsically activity-driven effect. Overall, our results generalize the correlated two-particle escape mechanism to interacting many-body active systems and establish particle density as a key control parameter governing the crossover from collective nonequilibrium barrier crossing to equilibrium-like independent escape.

Overall, our study establishes a unified picture of confined active matter in double-well potentials: small, highly persistent systems exhibit strong nonequilibrium correlations, collective synchronization, and enhanced fluctuations, while large, crowded systems transition toward equilibrium-like behavior with suppressed fluctuations and local dynamics. These findings highlight how confinement geometry, persistence time, and particle number can be tuned to regulate the magnitude and spatial organization of fluctuations in active systems. This model is also relevant to experimentally accessible confined active-colloid systems. Active Janus particles can be confined in optical traps, and bistable optical potentials have been used to study active-particle barrier crossing. A quasi-one-dimensional double-well setup with interacting active colloids could therefore directly test our predicted crossover from extended correlated barrier crossing at low density to localized collective crossing under strong crowding. The framework developed here may be extended to higher dimensions, asymmetric potentials, and systems with hydrodynamic interactions, providing a pathway to control and design collective behaviors in confined active materials.

\appendix

\section{Details of simulations and parameters}
\label{appendix:parameters}

We perform direct simulations of Eq.\ (\ref{aoup}) and Eq.\ (\ref{passive}) using the Euler–Maruyama method \cite{rahman2020convergence} with a time step $dt = 10^{-3}$ over $10^7$ iterations. The parameters are set as follows: particle radius $R = 1$ (as all lengths are scaled by $R$), interaction strength $k = 95$, well separation parameter $a/R = 10$, friction coefficient $\gamma = 10$, and propulsion speed $v_0/(R/t_0) = 12.5$. The particles are initially distributed equally across both wells. For passive particles, we use a diffusion constant $D/(R^2/t_0) = 0.625$ in Eq.\ (\ref{passive}). Simulations are performed for $10^7$ timesteps to ensure relaxation to steady state. The mean density $\langle\rho(x)\rangle$ is obtained by time-averaging after discarding the first $10^4$ steps. 
 We use $W_0 = 5$ in the subsequent calculations unless stated otherwise.

\section*{Conflict of Interest Statement}
The authors declare that they have no known financial or personal conflicts of interest that could have appeared to influence the work reported in this manuscript.

\section*{Availability of data}
Data for this article are available on request – please contact the corresponding author.

\section*{Acknowledgements}
The author acknowledges financial support by Science
and Engineering Research Board (SERB), Government of
India (File SRG/2022/000598) and (File MTR/2022/000281).
The author also acknowledges financial support from the
Birla Institute of Technology and Science, Pilani (File
PLN/AD/2022-23/3). The author also thanks DST-FIST for
the computational resources provided to the Department of
Physics, BITS Pilani.

\section*{Ethical Statement}

During the preparation of this work, the author(s) used ChatGPT (OpenAI) to assist with phrasing and language refinement, and Grammarly for typographical and grammatical corrections. After using these tools, the author(s) reviewed and edited the content as needed and take full responsibility for the content and scientific integrity of the publication.

\bibliography{bibliography}

\end{document}